\documentclass[a4paper,dvips 14pt]{revtex4}
\usepackage{epsfig,graphics,graphicx,amsmath,amssymb}

\begin{document}
	\baselineskip=.6cm
\title{On the difference of time-integrated $CP$ asymmetries in  $D^0\rightarrow K^+K^-$ and $D^0\rightarrow\pi^+\pi^-$ decays: unparticle physics contribution}

\author{Hosein Bagheri}
  \author{Mohammadmahdi Ettefaghi}
    \author{Reza  Moazzemi}

\affiliation{Department of Physics, University of Qom, Ghadir Blvd., Qom 371614-6611, I.R. Iran}

\begin{abstract}
The LHCb Collaboration has recently measured the difference of time-integrated $CP$ asymmetry in $D^{0}\rightarrow K^{+}K^{-}$ and $D^{0}\rightarrow \pi^{+}\pi^{-}$ decays, more precisely. The reported value is $ \Delta A_{CP}=-0.10 \pm 0.08 (\text{stat}) \pm 0.03 (\text{syst})\%  $ which indicates no evidence for $CP$ violation. We consider the possible unparticle physics contribution in this quantity and by using the LCHb data try to constrain the parameter space of unparticle stuff.
\end{abstract}


\maketitle

\section{Introduction}\label{1}
The standard model (SM), being a very successful model in all tests carried out so far, has been completed with the discovery of the Higgs boson, the last major missing piece of the SM, during the first LHC run. However, some problems such as baryon asymmetry, dark matter, neutrino oscillation etc. call for new physics beyond the SM. Therefore, physicists are now looking for phenomena that do not conform the SM predictions. In particular, to have an appropriate mechanism for the explanation of baryon asymmetry, one expects deviations of the SM predictions for $CP$ violation. Until now, $CP$ violation signals have been observed in the quark sector of the SM {\it e}.{\it g}. in K mesons \cite{christenson1964evidence} and B mesons \cite{carter1981cp,abe2001observation} decays which are consistent with the SM predictions. The SM prediction of $CP$ violation for D mesons is very small, so that ``a measurement of $CP$ violation in D decays would be a signal of new physics". For instance, in singly Cabbibo suppressed (SCS)
decays the $CP$ asymmetry is estimated to be about  ${\cal O}(0.05-0.1)$\% \cite{bord}.
In addition, there exist large uncertainties in the theoretical estimations of parameters so that the interpretation of measurements in the context of the SM encounters with some ambiguities.  In fact, the c quark is not heavy enough to
apply heavy quark effective theory (like in B physics), on the other hand it is not light
enough to use chiral perturbation theory (like in kaon physics). Moreover, the rate of mixing in neutral charm mesons is extremely small. Hence, the indirect $CP$ violations ($CP$ violation through mixing and the interference of the mixing and decay) are negligible in charm processes. In particular, when the difference of time-integrated $CP$ asymmetries in $D^0\rightarrow K^+K^-$ and $D^0\rightarrow\pi^+\pi^-$ decay is measured, the indirect contribution  almost vanishes in the difference.

The LHCb Collaboration has recently measured the difference between the $CP$ asymmetry in $D^{0}\rightarrow K^{+}K^{-}$ and $D^{0}\rightarrow \pi^{+}\pi^{-}$ to be $ \Delta A_{CP}\equiv A_{CP}(D^{0}\rightarrow K^{+}K^{-})-A_{CP}(D^{0}\rightarrow \pi^{+}\pi^{-})=[-0.10 \pm 0.08 (\text{stat}) \pm 0.03 (\text{syst})]\% $ \cite{prl2016}. This measurement has superseded the previous result obtained using the same decay channels \cite{LHCb1,CDF,Belle}.

According to the SM, the tree level amplitudes of $D^0\rightarrow K^+K^-$ and $D^0\rightarrow \pi^+\pi^-$ decays  involve only the first two quark generations, which cannot have the $CP$ violating Kobayashi-Maskawa (KM) phase. Therefore, both the weak and strong phases needed for the direct $CP$ violation come from the loop-induced gluon penguin diagrams. % in the case of singly-Cabibbo suppressed $D^0$ decays.
This implies that the SM prediction is loop suppressed as well as CKM suppressed; parametrically we have ${\cal O}((\alpha_s/\pi)(V_{ub}V^*_{cb})/(V_{us}V^*_{cs}))\sim 10^{-4}$. Several authors have tried to improve estimates for $\Delta A_{CP}$ in the SM \cite{bord,sm2,sm3}. Although there is large uncertainty in the SM value of $\Delta A_{CP}$, it is important and exciting to consider the recent measurement of $\Delta A_{CP}$ as possible new physics and explore allowed parameter space \cite{np0,np1,np2}.

It has been conjectured that there may exist ``stuff" that does not necessarily have zero mass but is still scale-invariant. This stuff cannot be described as particle, so it has been called ``unparticle" \cite{georgi1}. If the unparticle stuff exists, it must couple with normal matter weakly, since there is not yet any observed signal confirming it. However, several new physical results due to the existence of unparticle have been explored extensively, see for instance  \cite{davoudiasl2007constraining,lewis2007cosmological,mcdonald2009cosmological,freitas2007astro,hannestad2007unparticle,deshpande2008long,boyanovsky2010oscillation}. The unparticle effects are also involved in the study of various decays, beyond the SM \cite{georgi2007another,bhattacharyya2007unraveling,chen2009constraints,wei2009interpretation,jplee}. In particular,
the CMS Collaboration  has been recently searched for dark matter and unparticles in events with a Z boson and missing transverse momentum in proton-proton collisions at $\sqrt{s} = 13$ TeV \cite{cms2017}. Also, the peculiar $CP$ conserving phases in unparticle propagators lead to a significant impact on $CP$ violation \cite{upcp1,upcp2,upcp3,upcp4,upcp5}. Moreover, since the unparticle couplings to quarks can be complex in general, some $CP$ violating phases, in addition to the SM weak $CP$ phase, can arise from these new couplings. More recently, we study the $CP$ violation in Cabibbo favored decays of D mesons via unparticle physics \cite{rusta}. In this letter, we study this effect on singly Cabibbo suppressed (SCS) charm mesons decays. Namely, we obtain the unparticle contributions in the difference $CP$ asymmetry of $D^0\rightarrow\pi^+\pi^-$ and  $D^0\rightarrow K^+K^-$. Hereby, using the experimental measurement of this difference, we shall study the relevant parameter space of unparticle physics.

In the next two sections we first briefly review the unparticle physics then consider its contribution to the time-integrated $CP$ asymmetries of  $D^0\rightarrow K^+K^-$ and $D^0\rightarrow\pi^+\pi^-$ decays. In the last section we summarize our discussion and conclusions.

\section{A brief review of unparticle physics}\label{sec1}

One can suppose that the very high energy physics contains a scale invariant sector with a nontrivial IR fixed point (Banks-Zaks theory) \cite{bz}. The properties and signatures of this sector are different from the SM particles, hence Georgi termed it ``unparticle" \cite{georgi1}. Unparticles can interact with the SM particles through the exchange of particles with a large mass scale $ M_{\cal U} $.  Below  this scale, one can write nonrenormalizable couplings involving both SM fields and Banks-Zaks (BZ) fields suppressed by powers of $ M_{\cal U} $ as follows:
\begin{equation}\label{eff}
\frac{1}{M^k_{{\cal U}}}{\cal O}_{\rm{SM}}{\cal O}_{\rm{BZ}},
\end{equation}
where ${\cal O}_{\rm{SM}}$  (${\cal O}_{\rm{BZ}} $) is an operator with mass dimension $d_{\rm{SM}}$ ($ d_{\rm{BZ}} $) built out of the SM (BZ) fields.  The scale invariance of BZ sector emerges in an energy scale $\Lambda_{\cal U}$ where the BZ operators match onto unparticle operators. Here we have a dimensional transmutation due to the renormalizable coupling of BZ operators. Therefore, the effective interaction between unparticle and SM operators below $\Lambda_{\cal U}$ can be written as follows:
\begin{equation}
\frac{1}{M^{k}_{{\cal U}}}{\cal O}_{sm}{\cal O}_{BZ} \xrightarrow{\text{lower  energy}} \frac{C_{{\cal U}} \Lambda^{d_{BZ}-d_{{\cal U}}}_{{\cal U}}}{M^{k}_{{\cal U}}}{\cal O}_{\rm{SM}}{\cal O}_{{\cal U}},
\end{equation}
where $d_{{\cal U}}$  is the scaling dimension of the unparticle operator ${\cal O_U}$, and the constant $C_{{\cal U}}$ is a coefficient function. The Lorentz structures of unparticle can be different, i.e. $ \cal O_U\equiv\cal O_{\cal U},O_{\cal U}^\mu,O_U^{\mu\nu}$. Couplings of these operators to all possible gauge invariant SM ones, with dimensions
less than or equal to 4 are listed in \cite{smunparticle}. Here, we consider the following effective interactions of scalar and vector unparticle operators with operators composed of quarks,
\begin{equation}\label{efin}
\frac{c_{\rm{v}}^{q^{\prime} q}}{\Lambda_{{\cal  U}}^{d_{{\cal  U}}-1}}\overline{q}^{\prime}\gamma_{\mu}(1-\gamma_{5})q{\cal O}^{\mu}_{{\cal  U}}+\frac{c_{\rm{s}}^{q^{\prime} q}}{\Lambda_{{\cal  U}}^{d_{{\cal  U}}}}\overline{q}^{\prime}\gamma_{\mu}(1-\gamma_{5})q\partial^{\mu}\cal O_{{\cal  U}},
\end{equation}
where $c^{q^{\prime} q}_{\rm{v}}$ and $c^{q^{\prime} q}_{\rm{s}}$ are the dimensionless couplings. Moreover, the propagator of a scalar (vector) unparticle is given by \cite{georgi1}
\begin{equation}\label{propagator}
\int d^{4}xe^{ip.x}\langle0\vert T({\cal O}^{(\mu)}_{{\cal  U}}(x){\cal O}^{(\nu)}_{{\cal  U}}(0))\vert0\rangle=\Delta^{\rm{s(v)}}_{{\cal  U}}(p^{2})e^{-i\phi_{{\cal  U}}},
\end{equation}
with
\begin{equation}
\Delta^{\rm{s}}_{{\cal U}}(p^{2})=\frac{A_{d_{{\cal U}}}}{2\sin (d_{{\cal U}}\pi)}\frac{1}{(p^{2}+i\epsilon)^{2-d_{{\cal U}}}},
\end{equation}
\begin{equation}
\Delta^{\rm{v}}_{{\cal U}}(p^{2})=\frac{A_{d_{{\cal U}}}}{2\sin (d_{{\cal U}}\pi)}\frac{-g^{\mu\nu}+p^{\mu}p^{\nu}/p^{2}}{(p^{2}+i\epsilon)^{2-d_{{\cal U}}}},
\end{equation}
where $ \phi_{{\cal U}}=(2-d_{{\cal U}})\pi$ and
\begin{equation}
A_{d_{{\cal U}}}=\frac{16\pi^{5/2}}{(2\pi)^{2d_{{\cal U}}}}\frac{\Gamma(d_{{\cal U}}+1/2)}{\Gamma(d_{{\cal U}}-1)\Gamma(2d_{\cal U})}.
\end{equation}
Here, the transverse condition $ \partial_{\mu}{\cal O}^{(\mu)}_{\cal U}=0 $ is used and the phase factor in Eq. (\ref{propagator}) arises from
$ (-1)^{d_{{\cal U}}-2}= e^{-i\pi(d_{\cal U}-2)} $.

%%%%%%%%%%%%%%%%%%%%%%%%%%%%%%%%%%%%%%%%%%%%%%%%%%%%%%%%%%%%%%%%%%%%%%%%%%%%%%%%%%%%%%%%%%%%%%%%%%%%%%%%%%%%%%%%%%%%%%%%%%%%%%%%%
\section{Time-integrated $CP$ asymmetries of  $D^0\rightarrow K^+K^-$ and $D^0\rightarrow\pi^+\pi^-$ decays with unparticle}

The SM predicts very small $CP$ violation effects in $D$-meson decays because, with an excellent approximation, only the first two quark generations are involved. The $CP$ violation in neutral D decays could be direct or indirect (or both) which come from  di-penguin and box diagrams being very small \cite{cfsm}.
The time-dependent $CP$ asymmetry of $D^0$ decay to final $CP$ eigenstates $f=K^+K^-,\pi^+\pi^-$  can
be approximated as \cite{cdf,lhcb2017}
\begin{equation}
A_{CP}=\frac{\Gamma(D^0(t)\rightarrow f)-\Gamma(\bar{D}^0(t)\rightarrow f)}{\Gamma(D^0(t)\rightarrow f)+\Gamma(\bar{D}^0(t)\rightarrow f)}\approx a^{\rm{dir}}_{CP}-A_\Gamma\frac{t}{\tau_D},
\end{equation}
where
\begin{equation}\label{adir}
a^{\rm{dir}}_{CP}=A_{CP}(t=0)=\frac{\Gamma(D^0\rightarrow f)-\Gamma(\bar{D}^0\rightarrow f)}{\Gamma(D^0\rightarrow f)+\Gamma(\bar{D}^0\rightarrow f)}
\end{equation}
and $\tau_D$ is
the average lifetime of the $D^0$ decay. Here, $A_\Gamma$ is the asymmetry between the $D^0$ and $\bar{D}^0$ effective decay widths. The indirect contributions in $D^0$ decays, $CP$ violation in mixing, are universal and negligible, hence $A_\Gamma$ is mostly related to the decay. It is independent of final states and measured to be about $ 10^{-4}$ \cite{lhcb2017}. Consequently, we can  write  the $\Delta A_{CP}$ as
\begin{equation}
\Delta A_{CP}\approx a^{\rm{dir}}_{CP}(K^+K^-)-a^{\rm{dir}}_{CP}(\pi^+\pi^-).
\end{equation}
 \begin{figure}[ht]
 	\centerline{\includegraphics[scale=0.45]{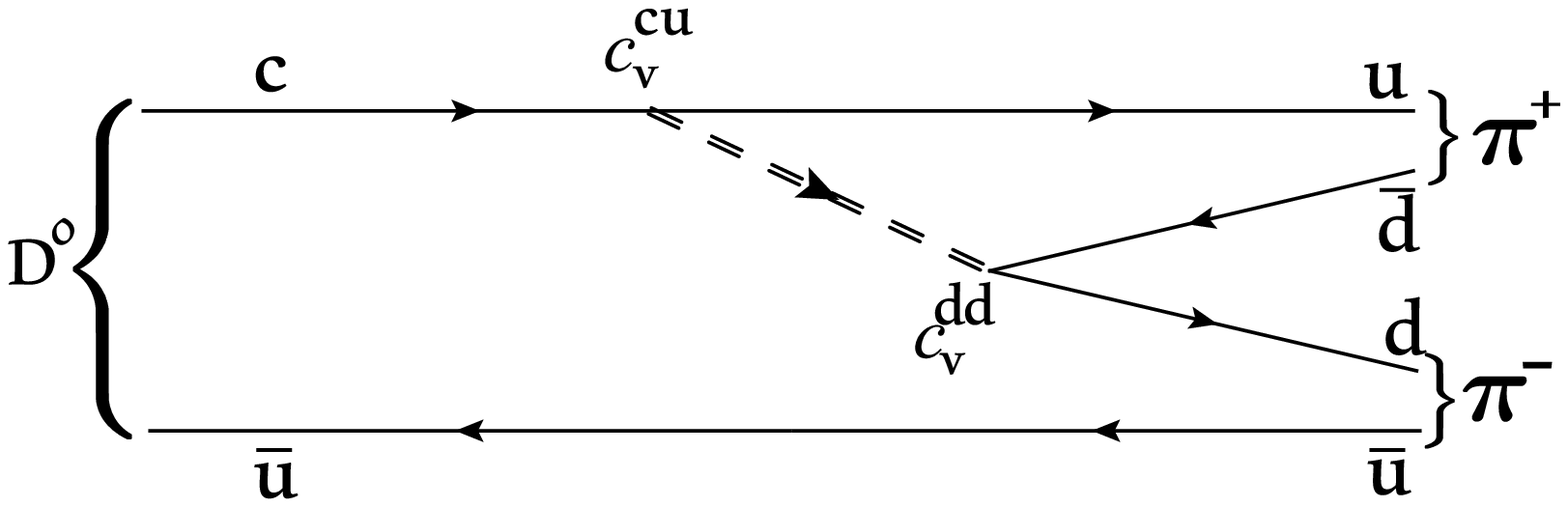} \includegraphics[scale=0.45]{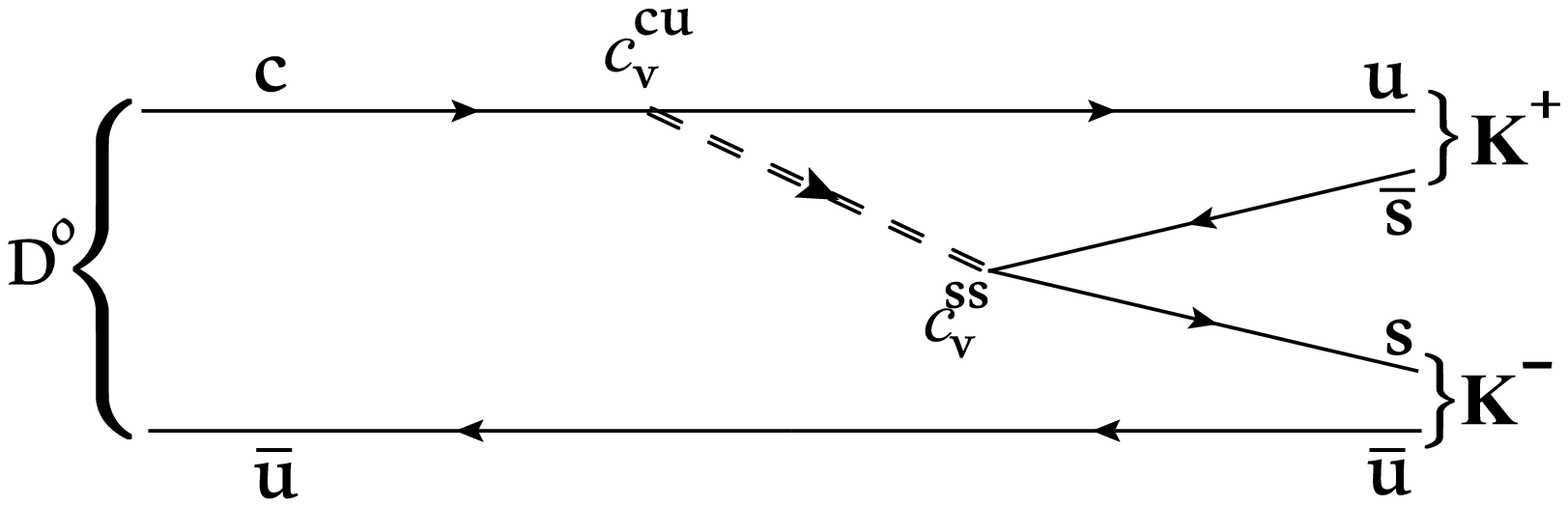}}
 	\caption{Diagrams for decay of $D^0$ to $\pi^+\pi^-$ and $K^+K^-$ final states via unparticle mediator.}\label{ddpk}
 \end{figure}
 Considering neutral unparticle such as Eq. (\ref{efin}) in addition to the SM contents (figure \ref{ddpk}), one leads to write the total amplitude, ${\cal A}_{\rm{tot.}}$, of the process $D\rightarrow\pi\pi(KK)$ as follows:
\begin{equation}
{\cal A}_{\rm{tot.}}= {\cal A}_{\rm{SM}} + {\cal A}_{\cal U},%= A_{SM}\left(1+r_{k^-\pi^+} \ e^{-i\phi_{\cal{U}}} e^{-i\phi_{W}} \right) \,,
\end{equation}
where ${\cal A}_{\rm{SM}}$ and $\cal A_{\cal U}$ are the SM and unparticle contributions, respectively. At the tree level, ${\cal A}_{\rm{SM}}$ is given by

 \begin{equation}
{\cal A}_{\rm{SM}}^{\pi\pi(KK)} = \dfrac{G_F}{\sqrt{2}}  V_{cd(s)}^*V_{ud(s)}  {\cal F}_{\pi(K)}\,,
 \end{equation}
where ${\cal F}_{\pi(K)}$ is a function which depends on the meson mass and QCD detail of the process, which will be finally removed in Eq. \ref{adir}. The unparticle contribution in amplitude is

\begin{equation}
{\cal A}_{\cal U}^{\pi\pi(KK)}=|{\cal A}_{\rm{SM}}^{\pi\pi(KK)}|\chi_{_{\pi(K)}}e^{-i\phi_{\cal U}}e^{-i\gamma_{\pi(K)}},
\end{equation}
 with
\begin{equation}
\chi_{_{\pi(K)}} =\dfrac{8 \, |c_{\rm{v}}^{dd(ss)} c_{\rm{v}}^{cu}| A_{d_{\cal{U}}} m_{W} ^2 }{2 \ g^2 a_1 N_c |V_{cd(s)}V_{ud(s)}| \sin({d_{\cal{U}}} \pi ) p^2} \left(\dfrac{p^2}{{\Lambda_{\cal{U}}}^2 }\right)^{d_{\cal{U}} -1}\,,
\end{equation}
where $a_1=C_2+C_1/N_C$ is the effective Wilson coefficient \cite{Buchalla}, $N_C$ is the color number and $ p^2 \sim m_D \bar{\Lambda}$ with  $\bar{\Lambda} =m_D-m_{\rm{c}}$. Here, $\gamma_{\pi(K)}$ is the weak phase related to the complex nature (in general) of the unparticle coupling coefficients, $c^{q^{\prime} q}_{\rm{v}}$.  Furthermore, $\phi_{\cal{U}} = d_{\cal{U}} \pi $ is the unparticle phase which plays the role of strong phase in the corresponding direct $CP$ violation. Note that, we have ignored the
 scalar unparticle contributions, since they  are suppressed by $m_D^2
 /\Lambda_{\cal U}^2$. Consequently,  the difference of time-integrated $CP$ asymmetry in  $D^0\rightarrow K^+K^-$ and $D^0\rightarrow\pi^+\pi^-$ decays from Eq. (\ref{adir}) becomes

\begin{equation}\label{deltaacp}
{\Delta A}_{CP} = \dfrac{2 \chi_{\pi} \sin({d_{\cal{U}}\pi}) \sin{\gamma_{\pi}}}{1+\chi_{\pi}^2+2 \chi_{\pi} \cos{d_{\cal{U}}\pi} \cos{\gamma_{\pi}}}-\dfrac{2 \chi_{K} \sin({d_{\cal{U}}\pi}) \sin{\gamma_{K}}}{1+\chi_{K}^2+2 \chi_{K} \cos{d_{\cal{U}}\pi} \cos{\gamma_{K}}} \,.
\end{equation}
 \begin{figure}[ht]
 	\centerline{\includegraphics[scale=0.75]{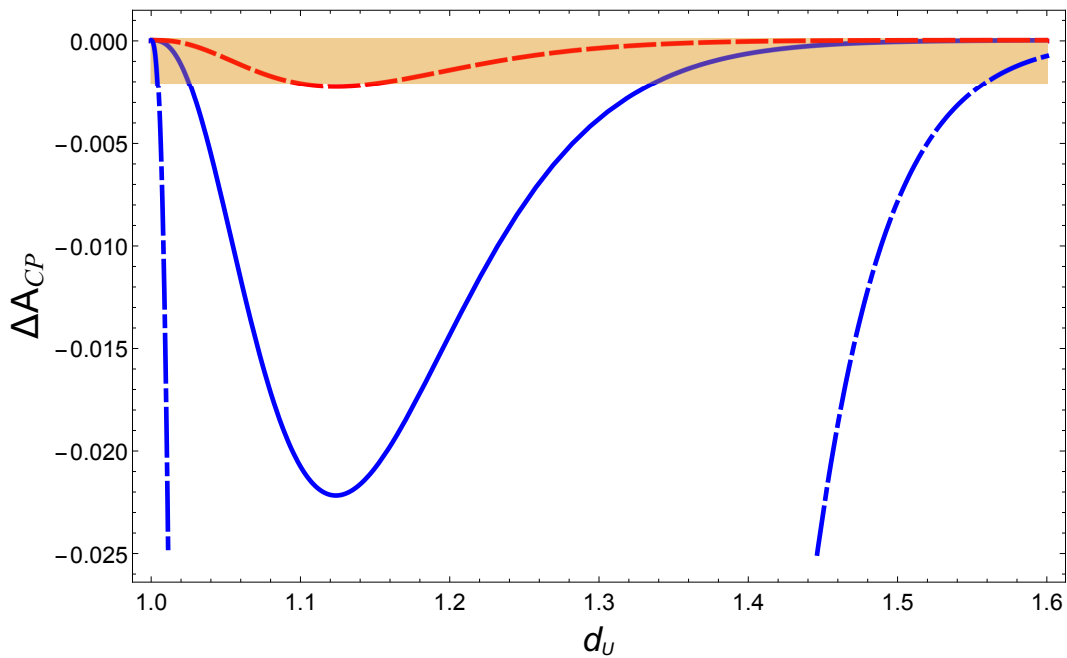}}
 	\caption{$\Delta A_{CP}$ in terms of the scaling dimension $d_{\cal U}$ by fixing weak phases to $\gamma_\pi=-1$, $\gamma_K=2.14$ and couplings to $|c_{\rm{v}}^{dd(ss)} c_{\rm{v}}^{cu}|\sim10^{-4}$ (dot-dashed),$10^{-6}$ (solid) and $10^{-7}$ (dashed), for the scale of unparticle $\Lambda_{\cal U}=15$ TeV.  The dark region shows the LHCb bounds.}\label{fig1}
 \end{figure}
 \begin{figure}[ht]
 	\centerline{\includegraphics[scale=0.75]{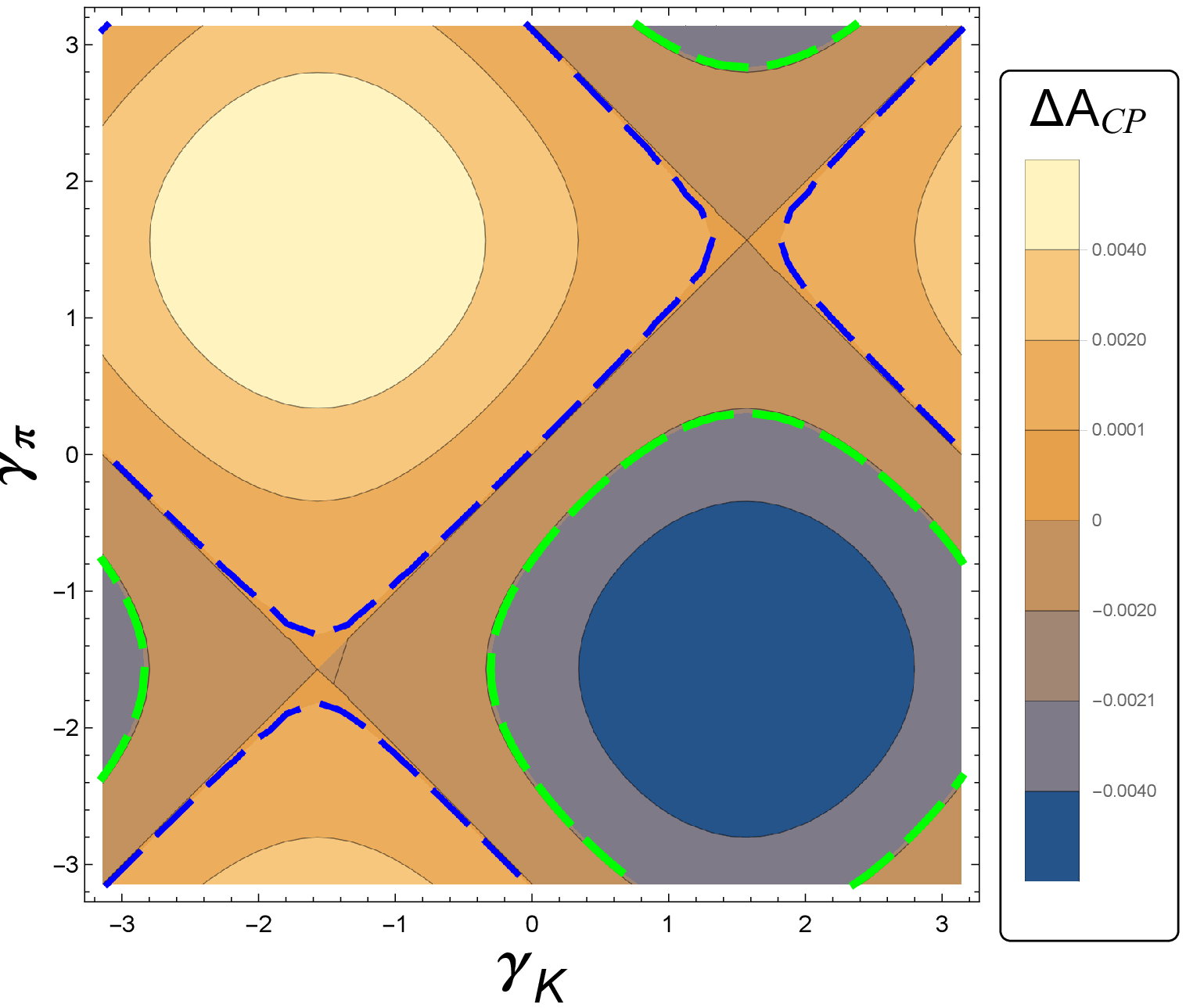}}
 	\caption{Contourplot of $\Delta A_{CP}$ in terms of the weak phases with scaling dimension $d_{\cal U}=1.85$ and unity couplings. Dashed lines correspond to LHCb upper (blue) and lower (green) bounds. The region between dashed lines is allowed and otherwise excluded.}\label{fig2}
 \end{figure}
We try to explore the relevant parameter space through some diagrams. We take the scale of unparticle $\sim 15$ TeV. First, the role of scaling dimension $d_{\cal U}$, which determines the strong phase, is illustrated in Fig. \ref{fig1} for three different product of couplings s and fixed values of the weak phases. In this figure, we take $\gamma_\pi=-1$, $\gamma_K=2.14$, which are corresponding to the  approximate maximum of $|\Delta A_{CP}|$ (see Fig. \ref{fig2}), and $|c_{\rm{v}}^{dd(ss)} c_{\rm{v}}^{cu}|\sim10^{-4},10^{-6},10^{-7}$. From this figure it is obvious that,  with the current precisions of measurements, there is no exclusion region for $|c_{\rm{v}}^{dd(ss)} c_{\rm{v}}^{cu}|\lesssim 10^{-7}$. It is noticeable that,  in the case of $10^{-4}$, the vertex factor in Eq. (\ref{efin}) (the ratio of a coupling and $\Lambda_{\cal U}^{d_{{\cal  U}}-1}$) is comparable to the least one chosen in \cite{upcp1}. In particular, note that while the unparticle has not been excluded by $B_{\rm{d}}\to \pi^+\pi^-$ for this choice, here it is seriously excluded  for $d_{\cal U}$ smaller than about 1.56. Second, in Fig. \ref{fig2}, $\Delta A_{CP}$ is plotted in terms of weak phases, $\gamma_\pi$ and $\gamma_K$ for a fixed value of $d_{\cal U}=1.85$ and unity couplings. We find that, if two terms in Eq. (\ref{deltaacp}) are in opposite phases, they are summed constructively. In this figure, the region between dashed lines is allowed and otherwise excluded. Here we should mention that, the future precise experiments, such as Phase-I and Phase-II upgrade of LHCb, or future analysis based on data collected of proton-proton collisions at the LHC, at a center-of-mass energy of  $13$ TeV, may lead to narrower allowed region, hence  one can constrain the parameter space more strongly.

 \section{Summary and conclusions}
 Recently, the difference of time-integrated $CP$ asymmetry in
  $D^0\rightarrow K^+K^-$ and $D^0\rightarrow\pi^+\pi^-$ decays $\Delta A_{CP}$, is
   measured by LHCb more precisely, which shows no evidence for $CP$ violation. This can impose strong
    constraints on the parameter space of any proposed new physics.
     In this letter, we suppose unparticle physics to contribute in
      $\Delta A_{CP}$. Here, there are seven parameters, $c^{dd}, c^{ss}, c^{cu}, d_{\cal U}, \gamma_\pi,\gamma_K$ and
       $\Lambda_{\cal U}$ which form our parameter space. Taking $\Lambda_{\cal U}\sim 15$ TeV, we have tried to illustrate the role
     of  other parameters.  According to Fig. \ref{fig1}, for $|c_{\rm{v}}^{dd(ss)} c_{\rm{v}}^{cu}|\gtrsim 10^{-7}$ there is an excluded region, due to the present precisions. In addition, for $|c_{\rm{v}}^{dd(ss)} c_{\rm{v}}^{cu}|\sim 10^{-4}$ we do not see an allowed region for $d_{\cal U}\lesssim 1.56$, while for the same order of couplings, all $d_{\cal U}$'s are allowed with $B_{\rm{d}}\to \pi^+\pi^-$ \cite{upcp1}. Also, the dependence of $A_{CP}$
        on weak phases is shown via contourplot in Fig. \ref{fig2}.
        Considering the recent LHCb results, this figure illustrates the excluded regions for choosing of couplings $\sim1$ and  $d_{\cal U}=1.85$.

\end{document}